\documentclass{PoS}

\title{Search for a narrow baryonic state decaying to $pK_S^0$ and $\overline{p}K_S^0$ in deep inelastic scattering at HERA}

\ShortTitle{Search for a narrow baryonic state decaying to $pK_S^0$ and $\overline{p}K_S^0$ in deep inelastic scattering at HERA}

\author{\speaker{Ryuma Hori}%
        \thanks{on behalf of the ZEUS Collaboration.}\\
       KEK, High Enegry Accelerator Research Organization, Oho1-1, Tsukuba, Ibaraki, Japan\\
       E-mail: \email{ryuma@post.kek.jp}} 

\usepackage{url}
\usepackage{graphicx}
\usepackage{xspace}
\usepackage{hangcaption}
\usepackage{mcite}
\newcommand{\bvec}[1]{\mbox{\boldmath $#1$}}

\newcommand{\figref}[1]{Fig.~\ref{#1}}
\newcommand{\figsref}[1]{Figs.~\ref{#1}}

\newcommand{\unit}[2]{#1\,#2}

\newcommand{\MeV}{\rm{}\,MeV}
\newcommand{\GeV}{\rm{}\,GeV}

\newcommand{\SGeV}{{$\rm{}\,GeV^2$}}

\abstract{
This paper presents the result of a search for a narrow strange baryonic state in the invariant-mass distribution of $pK^0_S(\overline{p}K^0_S)$ in HERA\,II data (2003-2007). The search was performed in the deep inelastic scattering (DIS) data with photon virtuality, $Q^2$, between 20 and \unit{100}{\SGeV}. The analysis in HERA\,II period has improvements over the previous ZEUS analysis in HERA\,I period (1996-2000). These consist of large statistics and improved particle identification performance by using both of the CTD $dE/dx$ and the MVD $dE/dx$ in this search. 
Contrary to evidence for such an exotic state around \unit{1.52}{\GeV} in the HERA\,I data, no such state is found in this analysis. 
The upper limits on the production cross section are set at 95\% confidence level.
}

\FullConference{XXIV International Workshop on Deep-Inelastic Scattering and Related Subjects\\
		11-15 April, 2016\\
		DESY Hamburg, Germany}

\begin{document}

\section{Introduction}
\label{sec-int}
\noindent
The LEPS experiment reported the observation of a narrow baryon resonance around \unit{1.53}{\GeV} in \\ 2003~\mcite{prl:91:012002,pr:c79:025210}. Since this resonance was decaying into $nK^+$, it is exotic and is impossible to be explained with a three-quark state. It could be explained as a state of five quarks i.e.\ a pentaquark state and it is named as $\Theta^+$($uudd\overline{s})$. The $\Theta^+$ state has been searched with the decay modes $nK^+$ or $pK^0_S (\overline{p}K^0_S)$ in many experiments. Some experiments confirmed the state, but others could not find it.

\begin{figure}
\begin{center}
\begin{tabular}{cc}
\begin{minipage}{0.35\hsize}
\includegraphics[width=\hsize, clip, trim= 0 5 0 0]{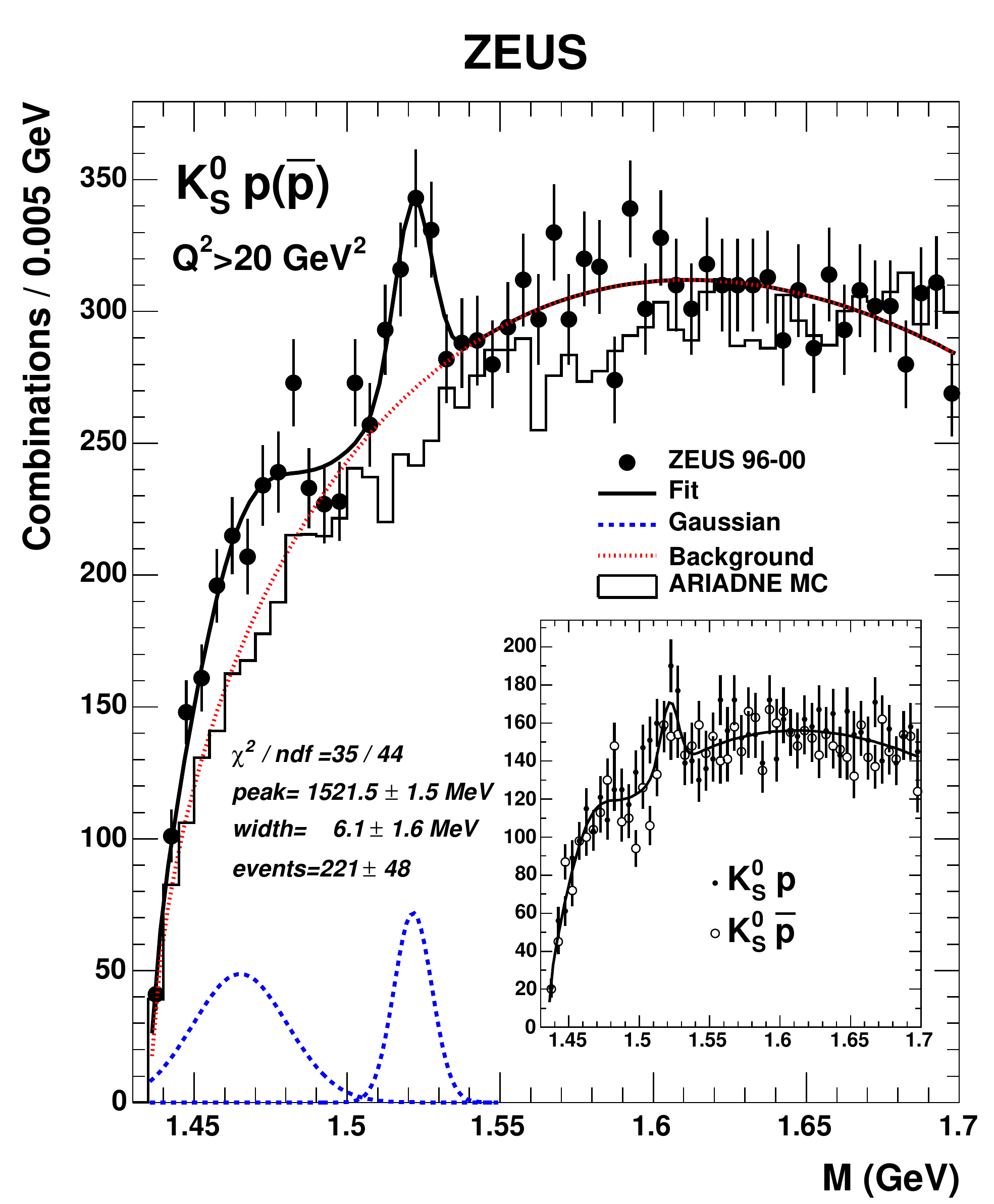}
\end{minipage}
\begin{minipage}{0.65\hsize}
\vspace{.5cm}
\caption{Invariant-mass spectrum for the $pK^0_S$ in DIS events with $Q^2$ $>$ $\unit{20}{\GeV^2}$ in HERA\,I analysis;
The solid line is the result of a fit using the background function plus two Gaussians. The dashed lines show the Gaussian components corresponding to the peak of the $\Theta^+$ signal and the $\Sigma(1480)$ bump, respectively. The inset shows the $pK^0_S$ (black dots)  and $\overline{p}K^0_S$ (open circles) candidates separately.} \label{HERA1result} 
\end{minipage}
\end{tabular}
\end{center}
\end{figure}

\noindent
As shown in \figref{HERA1result}, the ZEUS experiment reported an evidence of a peak structure, consistent with the $\Theta^+$, around \unit{1.52}{\GeV} in the $pK^0_S$ invariant mass distribution\footnote{Particle names refer to both particles and anti-particles, unless otherwise stated.} in electron-proton deep inelastic scattering (DIS) data with the range of photon virtuality $Q^2$ $>$ $\unit{20}{\GeV^2}$~\cite{pl:b591:7}. The data was collected in the HERA\,I period (1996--2000) and the integrated luminosity was $121$~pb$^{-1}$.
The H1 collaboration also searched for the $\Theta^+$ in DIS events with a similar kinematic region 20 $<$ $Q^2$ $<$ $\unit{100}{\GeV^2}$, but H1 did not find any peak and an upper limit was obtained~\cite{pl:b639:202}. However, this H1 limit did not exclude the ZEUS result.
To confirm or refute the $\Theta^+$ resonance, a search has been performed in the HERA\,II period (2003--2007) with a larger statistics ($358$~pb$^{-1}$) and with an improved ZEUS tracking system. A silicon-strip micro vertex detector (MVD) was installed closer to the beam line than the central tracking detector (CTD). It could provide more information of the ionization energy loss per unit length, $dE/dx$ and improve the efficiency of the proton particle identification (PID) from a huge background of mainly pions. 

\noindent
This paper presents the result~\cite{PQpaper} of a search for a narrow resonance in the $pK^0_S$ system of high-energy $ep$ collisions in the HERA\,II data.

\section{Event selection}
\subsection{Event sample}
\noindent
In the reconstruction of DIS events, the kinematic variables were calculated by the scattered electron and the remaining particles in the calorimeter (CAL). 
In order to compare with both of ZEUS and H1 HERA\,I analyses, the $Q^2$ range was set to between 20 and \unit{100}{\SGeV}. Other DIS kinematic selections were similar to those in the ZEUS HERA\,I analysis~\cite{pl:b591:7}. 
 
\subsection{$\bvec{K^0_S}$ and proton selection}
\noindent
Neutral strange mesons $K^0_S$ decaying into $\pi^+ \pi^-$ were reconstructed by a vertex constraint fit of two oppositely charged tracks~\cite{PQpaper}.  
Each track was required to pass through at least three inner superlayers of the CTD, to have at least three MVD hits, and to have transverse momentum $p_T > \unit{0.15}{\GeV}$ and pseudo-rapidity $|\eta| < 1.75$.
Geometrical selections for the $K^0_S$ vertex were performed by its position and co-linearity angles.
$K^0_S$ candidates were also requested to have kinematic ranges $p_T(K^0_S) > \unit{0.25}{\GeV}$ and $\mid \eta(K^0_S) \mid < 1.6$ and mass range $0.482$ $<$ $M(\pi^+\pi^-)$ $<$ $\unit{0.512}{\GeV}$.
In addition, two mass requirements were imposed to eliminate contamination from photon conversions and $\Lambda$ decays. 

\noindent
The proton tracks were selected by kinematic requirements and PID selection. 
The kinematic selec-\\
tions on proton candidates were: its momentum, $p_{\rm{track}}$, satisfies 0.2 $<$ $p_{\rm{track}}$ $<$ \unit{1.5}{\GeV}; it passes through at least three inner superlayers of the CTD, and has at least two MVD hits. 
The PID selection was performed by combining the CTD and MVD $dE/dx$ information~\cite{thesis}. 
Each $dE/dx$ resolution was $\approx$10\% and selections of each $dE/dx$ were treated similarly. 
First, protons were selected by requiring that the $dE/dx$ value is greater than 1.15 in units of minimum-ionizing particles (mips) and is within the band defined from the parameterized Bethe--Bloch function~\cite{chin:c38:090001}.
\begin{figure}[b]
\begin{center}
\begin{tabular}{cc}
\begin{minipage}{0.70\hsize}
\includegraphics[width=\hsize, clip, trim= 0 0 0 100]{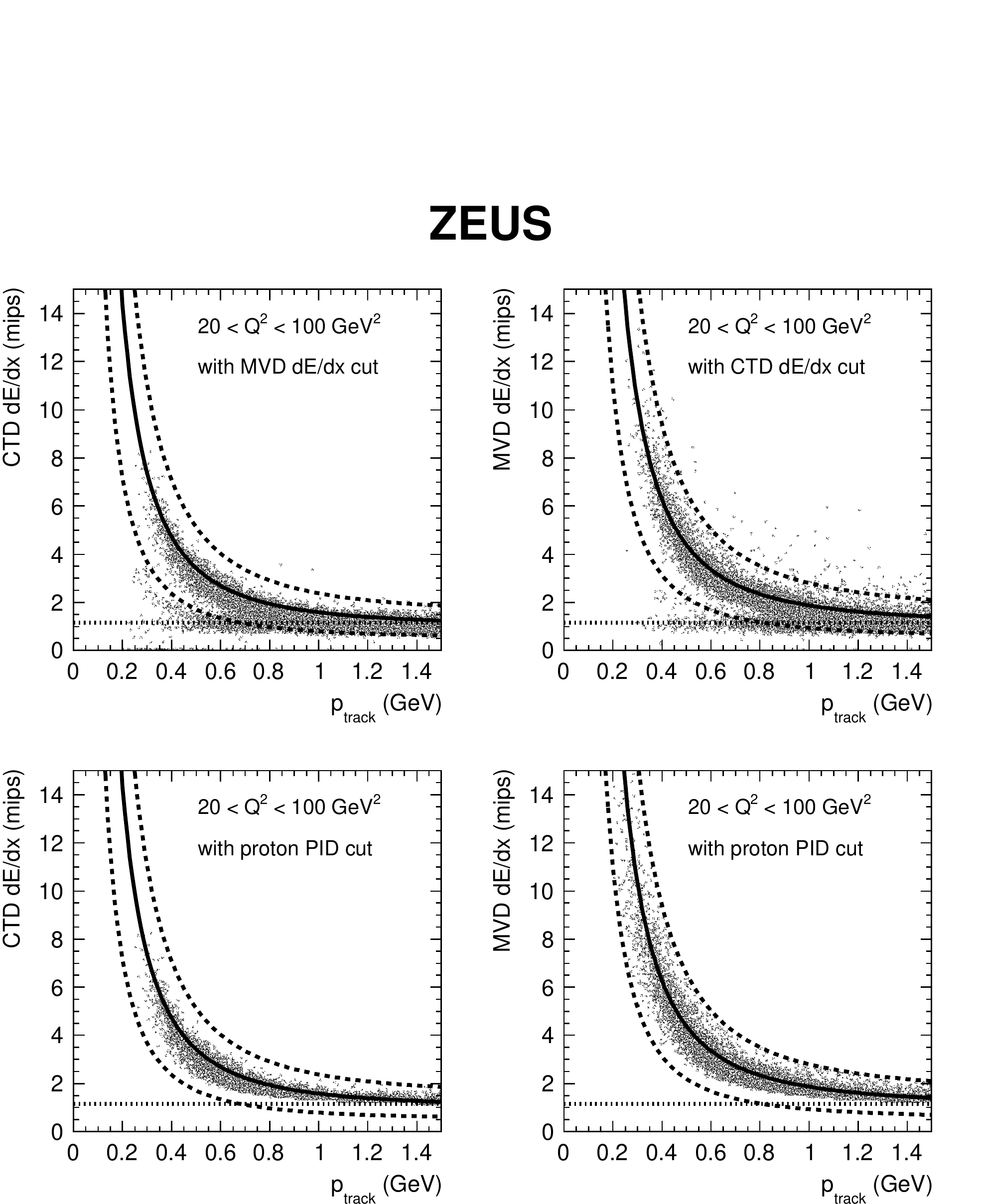}
\hspace*{50mm}
\\[-107.5mm]

\hspace{4.1cm}
(a)\hspace{4.85cm}(b)\\[4.1cm]

\hspace{4.1cm}
(c)\hspace{4.85cm}(d)\\[2.5cm]
\end{minipage}
\begin{minipage}{0.3\hsize}
\vspace{.5cm}
\caption{The $dE/dx$ vs. $p_{\rm{track}}$ distributions for (a) the CTD and (b) the MVD for proton candidates selected by the other detector. (c) the CTD and (d) the MVD for final proton candidates, including tracks with only MVD $dE/dx$ information. 
The solid lines show the predicted Bethe--Bloch function for the proton. The dashed lines indicate the limits used for bands of the proton selection. The dotted line shows the selection at $1.15$ mips.} \label{DISEV_P} 
\end{minipage}
\end{tabular}
\end{center}
\end{figure}
The CTD and MVD $dE/dx$ distributions selected as protons by the other detector are shown in \figsref{DISEV_P} (a) and (b), respectively. The clear proton bands are visible, but the contaminations from kaons and pions are still seen. 
As the next selection, a likelihood-like estimator was defined based on distances to the predicted Bethe-Bloch functions from each of the assumptions that the track is proton, kaon and pion.
If the CTD $dE/dx$ was not determined due to a signal saturation, protons were selected by only the MVD $dE/dx$. 
The CTD and MVD $dE/dx$ distributions with final proton PID selection are shown in \figref{DISEV_P}(c) and (d), respectively. The remarkable contaminations from kaon and pion are not visible. 
The proton PID efficiency was estimated by using protons from a $\Lambda$-decay sample~\cite{thesis}.
The efficiency of the proton PID integrated over $p_{\rm{}track}$ between 0.1 and \unit{1.5}{\GeV} is 54\%. The pion-rejection factor was also calculated using pions from $K^0_S$ decays sample. The factor is above 1000 with $p_{\rm{}track}$ < \unit{1.2}{\GeV} and decreases to 100 at \unit{1.5}{\GeV}.
For a direct comparison with the HERA\,I analysis, another analysis was performed with the proton PID consisting of the $dE/dx$ band and the mip selections of the CTD $dE/dx$ only.
This results in a higher proton PID efficiency of 82\%, but the pion rejection factor with $p_{\rm{}track}$ $>$ \unit{0.6}{\GeV} is 10--100 times worse.
\section{Results}
\subsection{The $\bvec{pK^0_S}$ invariant-mass distribution} 
\noindent
Candidates of the $pK^0_S$ system were obtained by combining proton and $K^0_S$ candidates as selected above in the kinematic range $0.5 < p_T(pK^0_S) < \unit{3.0}{\GeV}$ and $|\eta(pK^0_S)| < 1.5$.

\begin{figure}
\begin{center}
\begin{tabular}{cc}
\begin{minipage}{0.35\hsize}
\includegraphics[width=\hsize, clip, trim= 270 270 0 140]{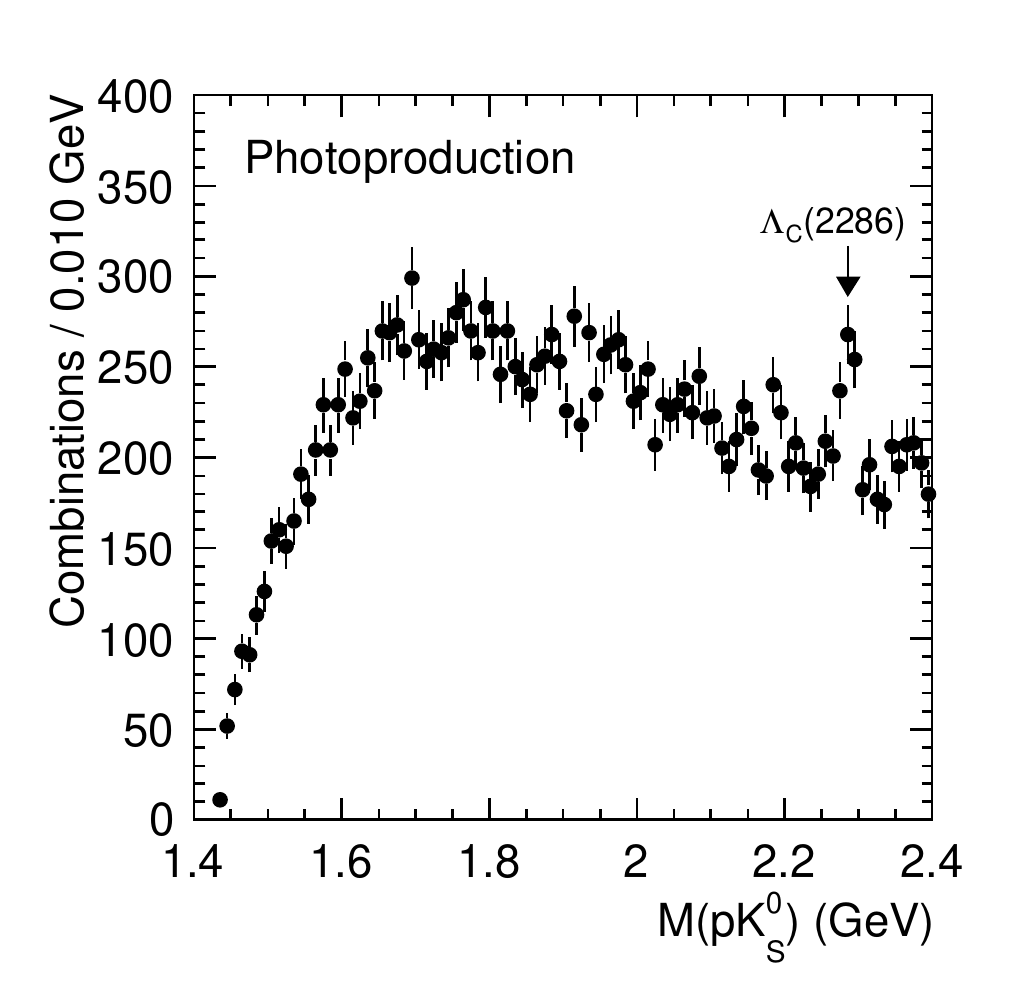}
\end{minipage}
\begin{minipage}{0.65\hsize}
\vspace{.5cm}
\caption{The $pK^0_S$ invariant-mass distribution in the mass range between 1.435 and \unit{2.4}{\GeV} in the photoproduction sample.} \label{DISEV_MASS_ph} 
\end{minipage}
\end{tabular}
\end{center}
\end{figure}

\noindent
In order to check the sensitivity to resonance searches in the HERA\,II data, the well-known $\Lambda_c(2286)$ baryon, having the same decay mode as the $\Theta^+$, was searched for in the $pK_S^0$ mass spectrum in a photoproduction sample. The photoproduction events have $Q^2 \approx \unit{0}{\GeV^2}$ and were collected by various event triggers requiring that no identified scattered electron was found in the CAL~\cite{thesis}.
A special cut of $p_T(pK^0_S)$ $>$ $0.15 \, E_{T}^{\rm{}\theta>10^\circ}$ was imposed to the photoproduction sample, where $E_{T}^{\rm{}\theta>10^\circ}$ is the sum of the transverse energy of the CAL cells outside a 10 degree cone from the proton-beam direction. This cut can enhance the signal of charmed particles.
A clear $\Lambda_c(2286)$ peak is seen in the $pK^0_S$ invariant-mass distribution in the photoproduction sample as shown in \figref{DISEV_MASS_ph}. 
The width of the $\Lambda_c$ peak is \unit{10}{\MeV}. This value is consistent with the width from MC simulation~\cite{thesis}. 

\begin{figure}
\begin{center}
\begin{tabular}{cc}
\begin{minipage}{0.7\hsize}
\includegraphics[width=\hsize, clip, trim= 0 0 0 420]{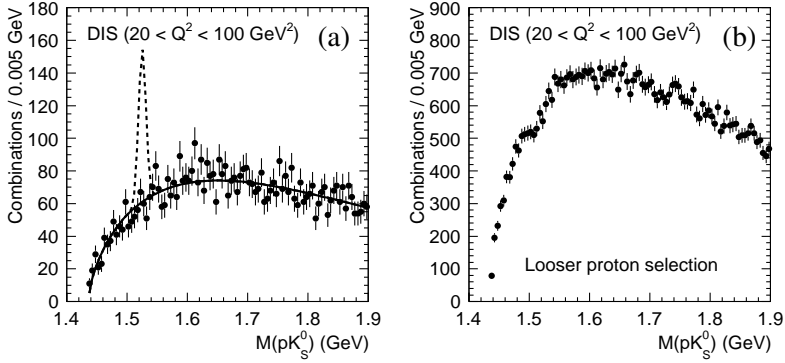}
\hspace*{5cm}
\\[-57.5mm]

\hspace{4.0cm}
(a)\hspace{4.9cm}(b)
\\[3.0cm]
\end{minipage}
\begin{minipage}{0.3\hsize}
\vspace{.5cm}
\caption{The $pK^0_S$ invariant-mass distribution in the DIS sample: (a) the distribution with the proton PID by both of the CTD and MVD. The solid line is the fitting result of the background function. The dashed line shows the signal estimated by the ZEUS HERA\,I result. (b) the distribution with PID from CTD only, as in the HERA\,I analysis.} \label{DISEV_MASS} 
\end{minipage}
\end{tabular}
\end{center}
\end{figure}

\noindent
The $pK^0_S$ invariant-mass distribution in the mass range between 1.435 and \unit{1.9}{\GeV} in the DIS sample is shown in \figref{DISEV_MASS}(a) with a fitting result by a function which consists of a Gaussian function for a postulated signal and of an empirical function of the form $\alpha (M-M_0)^{\beta} (1+\gamma (M-M_0))$ for the background, 
where $\alpha$, $\beta$ and $\gamma$ are free parameters, $M$ is the reconstructed mass of $pK_S^0$ system, and $M_0$ is the sum of the nominal proton and $K^0$ masses of PDG~\cite{chin:c38:090001}. The pion contamination in the proton candidates was estimated to be less than 10\%.
No significant peak is seen in the $pK^0_S$ invariant-mass distribution. An estimated peak of the $\Theta^+$ with a strength as reported in the ZEUS HERA\,I result is shown as a dashed line above a solid curve that represents the background-only fit in \figref{DISEV_MASS}(a). It is scaled by the differences of event selections, detector efficiencies and luminosity.
\noindent
In order to compare directly between this analysis and the ZEUS HERA\,I analysis, another analysis was performed with kinematic selections similar to those in the HERA\,I analysis, and the CTD-only $dE/dx$ PID selections as explained above.
The result is shown in \figref{DISEV_MASS}(b) and no peak is seen.
For this mass distribution, the pion contamination was estimated to be more than 50\%. 

\subsection{Upper limits on the production cross section}
\noindent
Three options for the width of the Gaussian in the fit function were used. One option was to fix it to $\unit{6.1}{\MeV}$ which was the value measured in the ZEUS HERA\,I analysis~\cite{pl:b591:7}. In the other two options, the width was set to 2.8-5.2\MeV (the detector mass resolution) and 5.6-10.4\MeV (2 times of the detector mass resolution)~\cite{thesis}.
The upper limit on the cross section at the 95\% confidence level (CL) was determined by increasing the $\chi^2$ value of the fit by 2.71~\cite{chin:c38:090001} from the best fit\footnote{Since the negative region of the assumed number of signal is not considered, a zero signal of the best-fit gives a more conservative limit than 95\% CL.}. 

\noindent
Detector acceptances for the event selections were estimated from the Monte Carlo (MC) samples except for the proton PID efficiency~\cite{thesis}. 
The signals were simulated with RAPGAP v.3.1030~\cite{cpc:86:147} by replacing $\Sigma^{\pm}(1189)$ with $\Theta^\pm$. It was assumed that the $p_T$ and $\eta$ distributions of the resonance are similar to the $\Sigma^{\pm}(1189)$ and the resonance decays isotropically to $pK^0_S$. The event simulation 
 was processed by the GEANT 3.21-based~\cite{tech:cern-dd-ee-84-1} ZEUS detector- and trigger-simulation programs.
The acceptances depended strongly on the $(p_T,\eta)$ values of the $pK^0_S$ system. 
Systematic uncertainties on the cross section were evaluated for the following four components~\cite{thesis}: uncertainty in the event selection (10\%), the proton PID efficiency with $\pm 1 \sigma$ of the measurement uncertainty (3\%), uncertainty in the mass-dependent selection efficiency (negligible) and the model uncertainty on the $p_T$ distribution of a $pK^0_S$ resonance (about 20\% at higher masses).
In addition, the uncertainty on the luminosity measurement was 2\%. 
All uncertainties were added in quadrature. 

\begin{figure}
\begin{center}
\begin{tabular}{cc}
\begin{minipage}{0.7\hsize}
\includegraphics[width=\hsize,clip,trim= 0 0 0 50]{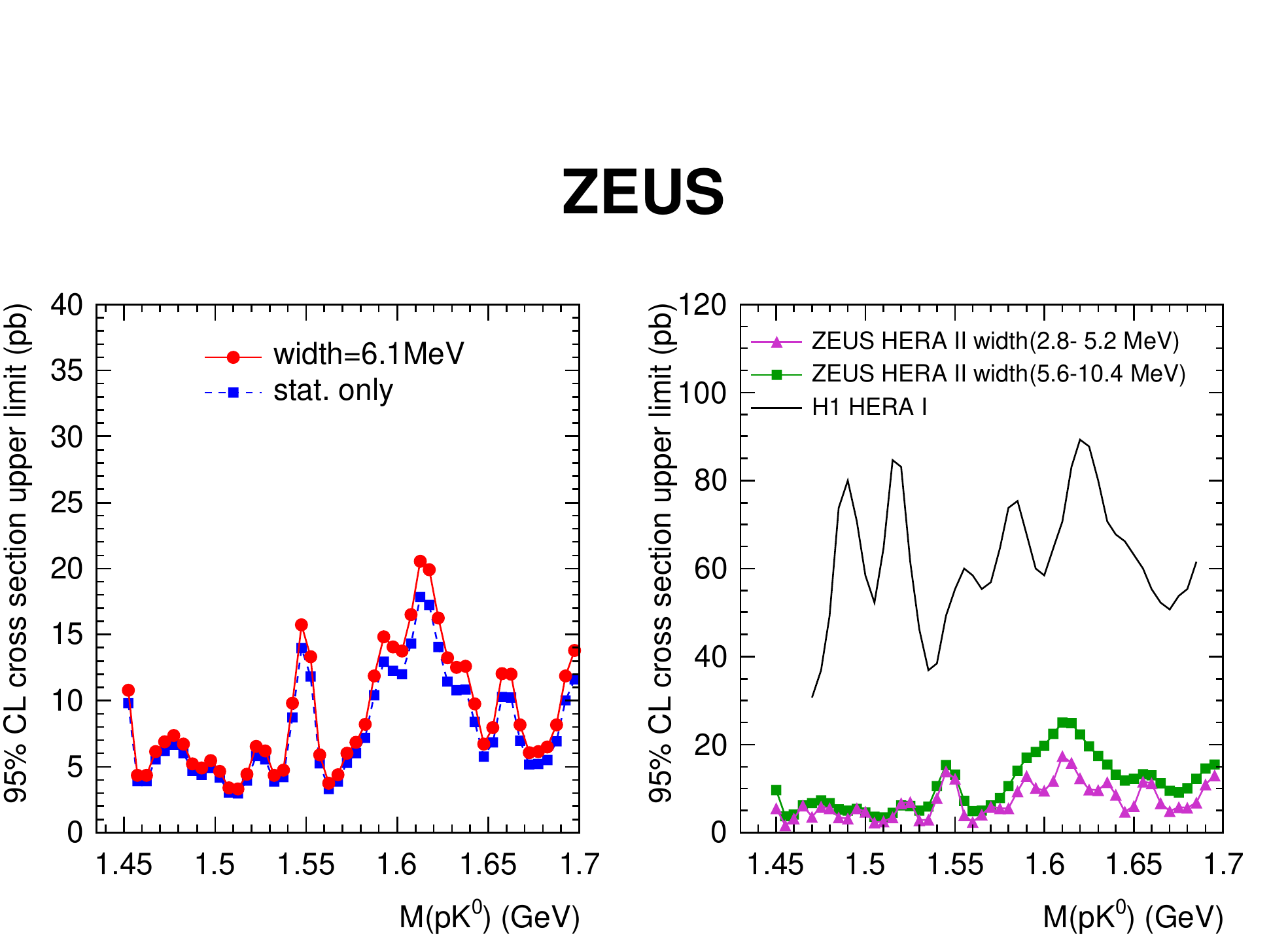}
\hspace*{50mm}
\\[-57.5mm]

\hspace{40mm}(a)\hspace{49mm}(b)
\\[30mm]
\end{minipage}
\begin{minipage}{0.3\hsize}
\vspace{5mm}
\caption{The 95\% CL upper limits of the $\Theta$ production cross section for different peak width conditions; 
(a) \unit{6.1}{\MeV}. The squares show the limit for only statistical error and the circles show the limit including systematic uncertainties. 
(b) the detector resolution and twice the detector resolution. The black line shows the limit from the H1 result.} \label{SUM_LIMIT}
\end{minipage}
\end{tabular}
\end{center}
\end{figure}

\noindent
The production cross section for either $\Theta^+$ or $\overline{\Theta^+}$, $\sigma ({\Theta})$, is defined as $(\sigma(ep \rightarrow e \Theta^+ X) + \sigma(ep \rightarrow e \overline{\Theta^+} X)) \times BR(\Theta^+ \rightarrow p K^0)$. The branching ratios used for the $K^0 \rightarrow K^0_S$ transition and for the $K^0_S \rightarrow \pi^+ \pi^-$ were 0.5 and 0.6895~\cite{chin:c38:090001}, respectively.
The results of determination of the upper limits\footnote{Since $K^0_S$ cannot be distinguished whether coming from $K^0$ or from $\overline{K^{0}}$ in this analysis, all limits also includes $p\overline{K^{0}}$ resonance.} of $\sigma ({\Theta})$ at 95\% CL are shown in \figref{SUM_LIMIT}(a) for the width of \unit{6.1}{\MeV} and in \figref{SUM_LIMIT}(b) for the width of the detector resolution and twice the detector resolution. 
The mass width of twice the detector resolution approximately corresponds to the width used in the calculation of the H1 HERA\,I result~\cite{pl:b639:202}. The ZEUS HERA\,II limit is smaller than the H1 limit.

\section{Summary}
\noindent
A search for a narrow strange baryonic state in the invariant-mass distribution of $pK^0_S(\overline{p}K^0_S)$ systems has been performed in the DIS data of the ZEUS HERA\,II period with an improved particle identification performance by using both of the CTD and MVD $dE/dx$. 
No evidence for a $\Theta^+$-like peak is found around \unit{1.52}{\GeV} and the result of the ZEUS HERA\,I analysis is not confirmed in this analysis.
The upper limits on the $\Theta$ production cross section with 3 different assumptions on the resonance width are set for the $pK^0$ mass range between 1.45 and \unit{1.7}{\GeV} in the kinematic ranges $0.5<p_T(pK^0)<\unit{3.0}{\GeV}$, $|\eta(pK^0)| < 1.5$ and $20 < Q^2 < $\unit{100}{\SGeV}.   

\providecommand{\urlprefix}{}
\providecommand{\etal}{et al.\xspace}
\providecommand{\coll}{Coll.\xspace}

\end{document}